\documentstyle[12pt,epsfig]{article}
\begin{document}
\def \ou{\alpha_1(\mu)}
\def \ot{\alpha_1(m_t)}
\def \tu{\alpha_2(\mu)}
\def \tt{\alpha_2(m_t)}
\def \thu{\alpha_3(\mu)}
\def \tht{\alpha_3(m_t)}
\def \uu{L^\dagger_U \dot L_U}
\def \dd{L^\dagger_D \dot L_D}
\def \ee{L^\dagger_E \dot L_E}
\def \smu{\hat {M}_U^2}
\def \smd{\hat {M}_D^2}
\def \sme{\hat {M}_E^2}
\def \syu{\hat {Y}_U^2}
\def \syd{\hat {Y}_D^2}
\def \sye{\hat {Y}_E^2}
\def \smf{\hat {M}_F^2}
\def \syf{\hat {Y}_F^2}
\def \tmf{\hat {M}_F}
\def \tmu{\hat {M}_U}
\def \tmd{\hat {M}_D}
\def \tme{\hat {M}_E}
\def \yuu{Y^{\dagger}_UY_U}
\def \ydd{Y^{\dagger}_DY_D}
\def \yee{Y^{\dagger}_EY_E}
\def \yff{Y^{\dagger}_FY_F}
\def \uuy{Y_UY^{\dagger}_U}
\def \ddy{Y_DY^{\dagger}_D}
\def \eey{Y_EY^{\dagger}_E}
\def \ffy{Y_FY^{\dagger}_F}

\def \cf{(1/16\pi^2)}
\def \cfi{16\pi^2}
\def \cu{2c_ig_i^2\hat {M}_U^2}
\def \cd{2c_i'g_i^2\hat {M}_D^2}
\def \ce{2c_i''g_i^2\hat {M}_E^2}
\def \cmu{[\hat {M}_U^2,L^{\dagger}_U\dot {L}_U]}
\def \cmd{[\hat {M}_D^2,L^{\dagger}_D\dot {L}_D]}
\def \cme{[\hat {M}_E^2,L^{\dagger}_E\dot {L}_E]}
\def \cvu{V\hat {Y}_D^2V^{\dagger}\hat {M}_U^2}
\def \cuv{\hat {M}_U^2V\hat {Y}_D^2V^{\dagger}}
\def \cvd{V^{\dagger}\hat {Y}_U^2V\hat {M}_D^2}
\def \cdv{\hat {M}_D^2V^{\dagger}\hat {Y}_U^2V}

\def \cft{m_t(m_t)A_u^{-1}}
\def \cfc{m_c(m_c){\eta_c}^{-1}A_u^{-1}}
\def \cfu{m_u(1GeV){\eta_u}^{-1}A_u^{-1}}
\def \cfta{m_{\tau}(m_{\tau}){\eta_{\tau}}^{-1}A_e^{-1}}

\def \tbt{\tan\beta(m_t)}
\def \tbb{\tan\overline{\beta}}
\def \vub{\overline{v}_u}
\def \vdb{\overline{v}_d}
\def \mtb{\overline{m}_t}
\def \mbb{\overline{m}_b}
\def \mab{\overline{m}_\tau}

\begin{flushright}
IC/99/10   \\
\end{flushright}

\vskip 50pt
\begin{center}
{\large \bf New Formulas and Predictions for Running Masses at Higher Scales in
MSSM} 
\end{center}

\vskip 35pt
\begin{center}
M.K.Parida \footnote{E-mail:mparida@nehus.ren.nic.in} \footnote{Permanent
Address:Physics Department,North-Eastern Hill University,Shillong 793022,India}
\footnote{Regular ICTP Associate}\\ 
The Abdus Salam International Centre for Theoretical Physics,Trieste,Italy\\
\vskip 15pt
B. Purkayastha\\
Physics Department,North-Eastern Hill University,Shillong 793022,India\\
\end{center}

\vskip 30pt

\begin{abstract}
Including contributions of scale-dependent vacuum expectation 
values of 
Higgs scalars ,we derive new one-loop formulas analytically for running 
quark-lepton masses at higher scales in the  MSSM .Apart from the 
gauge-coupling dependence of all masses being different from earlier formulas,the
third-generation-Yukawa-coupling effects are  absent in 
the masses of the first two generations.
While predicting the masses and $\tan\beta$,numerically, 
we also include two-loop effects.
\end{abstract} 

\newpage
One of the important objectives of current researches in High Energy Physics 
is to understand the masses and mixings of quarks and leptons in the context
of a unified theory of basic interactions.Apart from accounting for the well
known gauge hierarchy problem,the minimal supersymmetric standard model(MSSM)
remarkably exhibits the unification of gauge couplings of the standard model
(SM) at $M_U\simeq 2\times 10^{16}GeV$ consistent with the CERN-LEP data 
\cite{R1}.The knowledge
of running  particle masses is not only essential near the electro-weak
scale,but also  near the intermediate and GUT scales 
in order to test successes of models  based upon quark-lepton 
unification \cite{R2},infrared structure of Yukawa couplings \cite{R3},Yukawa 
textures for fermion masses \cite{R4},and predictive ansatz for fermion  
masses and mixings in $SO(10)$ \cite{R5}.For model building ,including
flavour symmetry to explain neutrino masses and mixings,it might be
necessary to know,how the masses of the first two generations behave with 
respect to the third generation Yukawa couplings.
The possibility that quark-lepton unification might exist at the GUT scale 
was strongly indicated \cite{R6}
where the effects of gauge couplings on running masses  were derived
analytically and the necessity of analytic formulas including Yukawa coupling
effects has been emphasized \cite {R7}.While the top-down 
approach predicts the particle masses and mixings in terms of GUT-scale 
parameters,the bottom-up approach predicts the running masses at higher 
scales in terms of experimentally determined values at low energies.In
particular ,the quark-lepton masses at higher scales ($\mu > m_t$) are 
predicted through one-loop formulas \cite{R8},
\begin{eqnarray}
m_t(\mu)=\cft e^{(6I_t+I_b)}\nonumber\\
m_c(\mu) = \cfc e^{3I_t},  m_u(\mu) = \cfu e^{3I_t}\nonumber\\  
m_b(\mu) = m_b(m_b){\eta_b}^{-1}A_d^{-1}e^{(I_t+6I_b+I_{\tau})}\nonumber\\
m_i(\mu) = m_i(1GeV){\eta_i}^{-1}A_d^{-1}e^{(3I_b+I_{\tau})},i=s,d \nonumber\\
m_\tau(\mu) = m_\tau(m_\tau){\eta_\tau}^{-1}A_e^{-1}e^{(3I_b+I_\tau)}\nonumber\\ 
 m_i(\mu) = m_i(m_i){\eta_i}^{-1}A_e^{-1}e^{(3I_b+I_\tau)},i={\mu},e    
\end{eqnarray}
where the Yukawa-coupling($y_f$) integrals are defined as, 
$I_f = \cf \int_{\ln m_t}^{\ln \mu} y_f^2(t)dt$, $f=t,b,\tau$;
 and
\begin{eqnarray}
 A_u&=& {(\ou/\ot)}^{13/198}{(\tu/\tt)}^{3/2}{(\thu/\tht)}^{-8/9}\nonumber\\
 A_d&=&{(\ou /\ot)}^{7/198}{(\tu/\tt)}^{3/2}{(\thu/\tht)}^{-8/9} \nonumber\\
 A_e&=&{(\ou /\ot)}^{3/22}{(\tu/\tt)}^{3/2}  
\end{eqnarray}
All running-masses occurring on the R.H.S. of eq.(1) are  
determined from experimental measurements using their 
respective relations to the pole masses. 
The parameters $\eta_{\alpha}(\alpha=u,c,d,s,b,e,\mu,e)$ are the 
QCD-QED rescaling factors \cite{R9,R10,R11}.       
Assuming SUSY breaking scale at $\mu=m_t$,these formulas are derived using 
analytic solutions to one-loop RGEs
of Yukawa-coupling eigen-values \cite{R8,R9,R10,R11} and the relation among  
mass matrix $M_a$,the Yukawa matrix $Y_a$,and the VEV $v_a$,
\begin{equation}
   M_a(\mu)=Y_a(\mu)v_a                                
\end{equation}
where $v_a=v_u=v_0sin\beta ,a=U$ 
and $v_a=v_d=v_0cos\beta,a=D,E$,with $v_0=174.0 GeV$.The running masses 
of all fermions of the first two generations are seen to depend upon
third-genaration-Yukawa-coupling integrals. 
In deriving these 
formulas,the only scale dependence that has been assumed is through 
the Yukawa couplings,$Y_a(\mu)$ , and not through the vacuum expectation 
values(VEVs),$v_a$,of eq.(4).
Numerical values of masses have also been  reported very recently totally 
neglecting the scale dependence of the VEVs\cite{R10}.
   Below the electroweak symmetry breaking scale,the VEVs do not depend upon 
the mass scales($\mu$) and the  running masses,related to their 
corresponding pole masses, are to be taken as the actual ansatz of  
the MSSM when the $\mu$-depedence of VEVs are ignored.
But ,above the electroweak-symmetry breaking scale,
the $\mu$-depedence of the VEVs
are well defined \cite {R12,R13},
through their RGEs and beta functions  upto two-loops,
\begin{eqnarray}
 \cfi(d\ln v_{u,d}/dt)&=&\beta_{v_{u,d}}+two-loops \nonumber\\
 \beta_{v_u}&=&3g_1^2/20+3g_2^2/4-3Tr(\uuy)          \nonumber\\
 \beta_{v_d}&=&3g_1^2/20+3g_2^2/4-Tr(3\ddy+\eey)
\end{eqnarray} 
The $\mu$-dependence of the VEV in the SM has been considered 
in the cotext of infrared fixed points\cite {R3} and ,very recently,to predict
CKM parameters and the top-quark mass at higher scales \cite {R14}.Here
we confine ourselves to the case of MSSM only. 
The RGEs for the mass matrices for $\mu>m_t$ are obtained in a 
straight-forward manner combining the corresponding RGEs for $Y_a$ and
$v_a$,
\begin{eqnarray}
  \cfi(dM_U/dt)&=&(-c_ig_i^2+3\uuy+\ddy)M_U\nonumber\\
  \cfi(dM_D/dt)&=&(-c_i'g_i^2+\uuy+3\ddy)M_D\nonumber\\
  \cfi(dM_E/dt)&=&(-c_i''g_i^2+\eey)M_E
\end{eqnarray}
where
$c_i=(43/60,9/4,16/3)$,$c_i'=(19/60,9/4,16/3)$, and $c_i''=(33/20,9/4,0)$.\\
Defining the diagonal mass matrices($\hat {M}_F$) and  Yukawa 
matrices($\hat {Y}_F$) through biunitary transformation and the CKM
matrix(V) as \cite {R9},\\
$\hat {M}_F=L^{\dagger}_FM_FR_F$,$V=L^{\dagger}_UL_D$,$\smf=
L^{\dagger}_FM_FM^{\dagger}_FL_F$,$\syf=L^{\dagger}_FY_FY^{\dagger}_FL_F$,
we derive  RGEs for $\smf$,
\begin{eqnarray}
 d\smu/dt &=&\cmu+\cf(-\cu+6\syu\smu+\cvu+\cuv)\nonumber\\
 d\smd/dt &=&\cmd+\cf(-\cd++6\syd\smd+\cvd+\cdv)\nonumber\\
 d\sme/dt &=&\cme+\cf(-\ce+6\sye\sme)
\end{eqnarray}
where the dot inside the commutator on the RHS denotes the derivative with 
respect to the variable $t=\ln {\mu}$.The diagonal elements of 
$L^\dagger_F\dot L_F$,($F=U,D,E$) are fixed to be zero in the usual manner 
\cite {R9} through diagonal phase multiplication.
The RGEs for the Yukawa and the CKM matrix elements
remain the same as before \cite {R8,R9,R15}.Now using the diagonal elements of 
both sides of  eqs.(6),the RGEs  for the mass eigen values are obtained by 
ignoring the Yukawa couplings of first two generations,
\begin{eqnarray}
 \cfi(dm_j/dt)&=&[-c_ig_i^2+|V_{jb}|^2y_b^2]m_j,j=u,c\nonumber\\
 \cfi(dm_t/dt)&=&[-c_ig_i^2+3y_t^2+|V_{tb}|^2y_b^2]m_t\nonumber\\
 \cfi(dm_j/dt)&=&[-c_i'g_i^2+|V_{tj}|^2y_t^2]m_j,j=d,s\nonumber\\
 \cfi(dm_b/dt)&=&[-c_i'g_i^2+3y_b^2+|V_{tb}|^2y_t^2]m_b\nonumber\\
 \cfi(dm_j/dt)&=&[-c_i''g_i^2]m_j,j=e,\mu\nonumber\\
 \cfi(dm_{\tau}/dt)&=&[-c_i''g_i^2+3y_{\tau}^2]m_{\tau}
\end{eqnarray}
Integrating these equations and using the corresponding low-energy values,
the new formulas are obtained in the small mixing limit as,
\begin{eqnarray}
 m_t(\mu)=m_t(m_t)B_u^{-1}e^{(3I_t+I_b)}\nonumber\\
 m_c(\mu)=m_c(m_c){\eta_c}^{-1}B_u^{-1}\nonumber\\
 m_u(\mu)=m_u(1GeV){\eta_u}^{-1}B_u^{-1}\nonumber\\
 m_b(\mu)=m_b(m_b){\eta_b}^{-1}B_d^{-1}e^{(I_t+3I_b)}\nonumber\\
 m_i(\mu)=m_i(1GeV){\eta_i}^{-1}B_d^{-1},i=d,s\nonumber\\
 m_{\tau}(\mu)=m_{\tau}(m_{\tau}){\eta_\tau}^{-1}B_e^{-1}e^{3I_\tau}\nonumber\\
 m_i(\mu)=m_i(m_i){\eta_i}^{-1}B_e^{-1},i=e,\mu
\end{eqnarray}
where
\begin{eqnarray}
B_u &=& {(\ou/\ot)}^{43/792}{(\tu/\tt)}^{9/8}{(\thu/\tht)}^{-8/9}\nonumber\\
B_d &=& {(\ou/\ot)}^{19/792}{(\tu/\tt)}^{9/8}{(\thu/\tht)}^{-8/9}\nonumber\\
B_e &=& {(\ou/\ot)}^{1/8}{(\tu/\tt)}^{9/8}
\end{eqnarray} 

\par
For $\tan\beta=v_u/v_d$,the RGE is  obtained from the difference of the 
beta functions, 
$\beta_{v_u}-\beta_{v_d}$,and the values at higher scales are given by the  
one-loop analytic formula,
\begin{equation}
\tan\beta(\mu)= \tan\beta(m_t)e^{(-3I_t+I_b+I_{\tau})}
\end{equation}
\par
It is clear from (8)-(9) that, 
compared with (1)-(2),the new formulas have very significant
differences with respect to their functional dependence on gauge and Yukawa
couplings in all
cases.Also
the masses of the first two-generations are found to be independent of third
generation Yukawa
couplings.While deriving one-loop formulas using the see-saw mechanism in SUSY
GUTs,it has been shown that the left-handed neutrino masses of the first two
generations have no additional dependence on Yukawa couplings except through
respective up-quark masses \cite {R16}.Combining the present result with that 
of ref. \cite {R16},it turns out that all fermion masses of first 
two generations of MSSM are independent of Yukawa couplings of the third 
generation.   
In view of the present results ,apart from modifying analytic
formulas,earlier numerical mass predictions including ref.\cite {R10}, where
$\mu$-dependence of VEVs have been ignored,are to be rescaled by $v_u(\mu)/v_0$
for
up-quark masses and by $v_d(\mu)/v_0$ for down-quark and charged lepton masses.
While estimating masses,VEVs,and $\tan\beta$
at higher scales, we have solved all relevant RGEs,including those of VEVs and
$\tan{\beta}(\mu)$ ,upto two-loops with the same
inputs at $\mu=m_t$ as in ref. \cite {R10}.We  find that the input value of
$m_t(m_t)=171\pm12$ GeV gives rise to the perturbative limit
$y_t(M_{GUT})\le 3.54$ at 
$\tan\beta(m_t)\ge1.74^{+.46}_{-.28}$. Due to running governed by the 
corresponding RGE at two-loop level,this limit at the GUT-scale turns out
to be $\tan\beta(M_{GUT})\ge.52^{+.14}_{-.10}$.
To our knowledge,this is the
first result in the literature ,showing that actual solutions to RGEs permit 
$\tan\beta(M_{GUT})(\equiv\tan\overline {\beta}) < 1$ near the perturbative
limit of $y_t(M_{GUT})$.This gives rise to the possibility of a perturbation 
expansion in terms of $\tan\overline{\beta}$ in this
region.We also observe the saturation of the perturbative limit
for the b-quark Yukawa coupling($\overline{y}_b$) for $\tan\beta(m_t)\simeq 61$, 
similar to \cite {R10}.We have checked that the one loop analytic solutions agree
with the
full two-loop numerical solutions within 5-7\% of accuracy except near the
perturbative limits,where the discrepancy increases further due to larger two-loop 
effects.

\par

In Table I ,we present the predictions for VEVs,$\tan\beta$ and
masses at two different scales:$\mu=10^9$GeV, and $\mu=2\times10^{16}GeV$
for the input $\tan\beta(m_t)=10$.Our solutions of RGEs yield  very
significatly different values of $v_u(\mu)$ than the assumed scale
independent one ,although $v_d(\mu)$ is not very significantly 
different, for such a $\tan\beta\approx 10$.This feature leads to quite different
up-quark masses,the most prominent  being $m_t(\mu)$.The running VEV of $v_u$
reduces the central value of $m_t(\mu)$ to nearly 72\%(57\%) at the
intermediate(GUT) scale.Similarly, $m_u(\mu)$ and $m_c(\mu)$ are reduced
to 80\%(67\%) and 79\%(66\%) ,respectively,at the intermediate(GUT)-scale
as compared to \cite {R10}.
As $v_d(\mu)$ is closer to the assumed scale-independent value for 
$\tan\beta\simeq 10$,all the
down quark and the charged lepton masses are closer to the values
obtained in ref. \cite {R10}. But , it is clear , that significant
differences will appear in these cases also from the computations based upon
scale-independent assumption,in the larger $\tan\beta$-region.  

\par

In Table II,we present GUT-scale predictions of VEVs,$\tan\beta$ and third
generation fermion masses ,denoted  with overbars,as a function
of different input values of $\tan\beta(m_t)$.The GUT-scale value of 
$m_t(M_{GUT})$ is found to reach nearly a minimum ,which is approximately
half of its perturbative-limiting value, for $\tan\beta\simeq 10$.After this 
minimum is
reached,$m_t(M_{GUT})$ increases slowly with increasing $\tan\beta$ ; but the
increase is faster for $\tan\beta\ge 50$.Similaly $m_b(M_{GUT})$ shows more 
than 10\% increase both for smaller(larger) values of 
$\tan\beta$ below(above) 
$2.0(40.0)$ as compared to its value at $\tan\beta=10$.Also the numerical
solution to RGE for $tan\beta(\mu)$ exhibits its
GUT-scale value
($\tan{\overline\beta}$) to be significantly less than the low energy input
except until the input  approaches the value of $\tan\beta(m_t)\simeq 61$
corresponding to the saturation of the perturbative limit of $y_b(M_{GUT})$.
In this region ,the 
GUT-scale value of $\tan{\overline\beta}$ exceeds the corresponding low
energy input as shown for the case of $\tan\beta(m_t)=60$.
  We conclude that inclusion of the effects of running VEVs yields 
completely new fermion mass fomulas with respect to their dependence
on gauge and Yukawa couplings.In view of the results presented here  and 
formulas for neutrino masses \cite {R16},all  running masses of first two
generations in the MSSM are independent of runnings of third-generation-
Yukawa couplings.This behavior of masses may be contrasted with that of Yukawa
couplings ,which, for the first two-generations,depend upon the couplings 
of the third.Numerical estimations yield very significantly different
values on masses at higher scales and provides interesting new informations 
on the GUT-scale values of $\tan\beta$.

\vskip 15pt
 
One of us(M.K.P.) thanks Professor Goran Senjanovic and Professor K.S.Babu for
useful discussions and  Professor S.Randjbar-Daemi for encouragement.The 
AS ICTP Associateship grants of M.K.P are gratefully acknowledged.

\newpage

\begin{table}
\caption{\bf Predictions of masses ,VEVs, and $\tan\beta(\mu)$ at two different 
scales in MSSM for  $\tan\beta(m_t)=10.0$ and other 
low-energy values same as in ref. \cite {R10}.}
\begin{tabular}{ccccc}
\hline
Parameter&This analysis&Ref.[10]&This analysis&Ref.[10]\\
&$\mu=10^9GeV$&&$\mu=2\times 10^{16}GeV$\\
\hline
$\tan\beta$&$7.973$&$10$&$6.912$&$10$\\
$v_u(GeV)$&$142.123$&$173.130$&$128.085$&$173.130$\\
$v_d(GeV)$&$17.815$&$17.312$&$18.534$&$17.312$\\
$m_t(GeV)$&$107.52$&$149^{+40}_{-26}$&$73.55$&$129^{+96}_{-40}$\\
$m_c(GeV)$&$.3373$&$.427^{+.035}_{-.038}$&$.2003$&$.302^{+.025}_{-.027}$\\
$m_u(MeV)$&$1.178$&$1.470^{+.26}_{-.28}$&$.7059$&$1.04^{+.19}_{-.20}$\\
$m_b(GeV)$&$1.580$&$1.60\pm.06$&$1.004$&$1.00\pm.04$\\
$m_s(GeV)$&$.0478$&$.0453^{+.0057}_{-.0063}$&$.0292$&$.0265^{+.0033}_{-.0037}$\\
$m_d(MeV)$&$2.4018$&$2.28^{+.29}_{-.32}$&$1.4632$&$1.33^{+.17}_{-.19}$\\
$m_{\tau}(GeV)$&$1.5177$&$1.4695^{+.0003}_{-.0002}$&$1.2566$&$1.1714\pm.0002$\\
$m_{\mu}(MeV)$&$89.088$&$86.217\pm.00028$&$73.6226$&$68.59813\pm.00022$\\
$m_e(MeV)$&$.422$&$.40850306$&$.3487$&$.32502032$\\
\hline
\end{tabular}
\end{table}
\newpage
\vskip 25pt

\begin{table}
\caption{\bf Predictions of VEVs,$\tan\beta$,and third generation fermion masses 
at the GUT-scale as
a function of input values of $\tan\beta(m_t)$ and other input masses same as in
ref.\cite {R10}.The GUT-scale values have been denoted with overbars.}
\begin{tabular}{ccccccc}
\hline
$\tbt$&$\tbb$&$\vub$&$\vdb$&$\mtb$&$\mbb$&$\mab$\\   
\hline
$1.75$&$.521$&$48.497$&$93.078$&$146.144$&$1.280$&$1.253$\\
$2$&$.963$&$80.87$&$83.90$&$100.962$&$1.116$&$1.252$\\
$5$&$3.35$&$123.13$&$36.74$&$75.138$&$1.008$&$1.252$\\
$10$&$6.910$&$128.08$&$18.54$&$73.556$&$1.004$&$1.256$\\
$20$&$14.18$&$128.75$&$9.079$&$73.92$&$1.022$&$1.2739$\\
$30$&$22.1$&$127.899$&$5.787$&$75.321$&$1.0613$&$1.3078$\\
$40$&$31.443$&$126.15$&$4.0119$&$77.869$&$1.1317$&$1.3662$\\
$50$&$44.476$&$123.05$&$2.766$&$82.768$&$1.274$&$1.484$\\
$60$&$80.60$&$116.08$&$1.440$&$98.103$&$1.820$&$1.924$\\
\hline
\end{tabular}
\end{table}

\end{document}